\documentclass[twocolumn,prb,superscriptaddress,longbibliography]{revtex4-2}
\usepackage{graphicx}
\usepackage{color}
\usepackage{amsmath}
\usepackage{enumitem}
\usepackage{amssymb}
\usepackage{hyperref}
\usepackage[normalem]{ulem}
\usepackage{adjustbox}

\hypersetup{
	colorlinks = true,
	linkcolor = blue,
	citecolor = blue,
	urlcolor  = blue,
}

\newcommand{\be}{\begin{equation}}
	\newcommand{\ee}{\end{equation}}

\newcommand{\bea}{\begin{eqnarray}}
	\newcommand{\eea}{\end{eqnarray}}

\renewcommand{\vec}[1]{{\boldsymbol #1}}
\renewcommand{\epsilon}{\varepsilon}

\def\nn{\nonumber\\}

\begin{document}
	\title{Size effects on atomic collapse in the dice lattice}
	\date{\today}
	
	\author{D. O. Oriekhov}
	\affiliation{Instituut-Lorentz, Universiteit Leiden, P.O. Box 9506, 2300 RA Leiden, The Netherlands}
	
	\author{S. O. Voronov}
	\affiliation{National Technical University of Ukraine ``Igor Sikorsky Kyiv Polytechnic Institute'', Beresteiskyi Ave. 37, 03056, Kyiv, Ukraine}
	
	\begin{abstract}
		We study the role of size effects on atomic collapse of charged impurity in the flat band system. The tight-binding simulations are made for the dice lattice with circular quantum dot shapes. It is shown that the mixing of in-gap edge states with bound states in impurity potential leads to increasing the critical charge value. This effect, together with enhancement of gap due to spatial quantization, makes it more difficult to observe the dive-into-continuum phenomenon in small quantum dots. At the same time, we show that if in-gap states are filled, the resonant tunneling to bound state in the impurity potential might occur at much smaller charge, which demonstrates non-monotonous dependence with the size of sample lattice. In addition, we study the possibility of creating supercritical localized potential well on different sublattices, and show that it is possible only on rim sites, but not on hub site. The predicted effects are expected to naturally occur in artificial flat band lattices.
	\end{abstract}
	\maketitle
	
\section{Introduction}

The phenomenon of atomic collapse was firstly discussed in connection with Dirac equation describing electrons near super-heavy nuclei \cite{Pomeranchuk1945,Zeldovich1972}. With the discovery of two-dimensional materials with Dirac cones in spectrum the atomic collapse again attracted attention \cite{Pereira2007,Fogler2007PRB,Shytov2007PRL_1,Shytov2007PRL_2}. A number of experiments \cite{Wang2013Science,Mao2016NaturePhys,Jiang2017NNano} were performed in graphene to analyze the effects that appear when the charge of impurity exceeds certain value and becomes supercritical. The experiments investigated impurities made as clusters of atoms and well as vacancy or induced by STM tip. 

The new generation of two-dimensional materials that host flat bands in addition to Dirac cones again posed a question on which effects the Coulomb impurity will produce. In Refs.\cite{Gorbar2019,Han2019,Pottelberge2020} it was shown by analytical and numerical calculations for dice lattice that the bound states decouple also from the flat band. In the case of infinite lattice model the flat band fully decomposes into a continuous spectrum in the field of Coulomb impurity because the pure Coulomb potential is long-range \cite{Gorbar2019,Pottelberge2020}.
Additional simulations on a finite lattice model with large sample size have shown that the bound states decoupled from flat band anticross with atomic collapse states coming from the upper dipsersive band \cite{Wang2022PRB}. 

The present study is motivated by the experiments with electronic lattices such as Lieb \cite{Slot2017} and honeycomb with s-p hybridization \cite{Broeke2021,Freeney2022}, which might serve as potential platform to realize also the dice lattice. The artificial lattices of such kind have approximately 10 times larger lattice constant that atomically-thin materials and thus much weaker electron-electron interactions. Thus, the effects of single-particle physics in external field might be easier to observe than in strongly-correlated flat band systems. To analyze the role of size effects on the possibility of observation of atomic collapse we consider quantum dots of circular shape made of dice lattice.

From the classification of the dice lattice terminations \cite{Oriekhov2018} it is known that particular boundary conditions lead to the gapless spectrum. In the case of gapped model \cite{Dey2020,Hao2022} the number of in-gap states is formed. In the circular shape of quantum dot all such terminations appear, thus leading to the existence of in-gap edge states. The electrons from such energy levels, if filled, may tunnel to the bound state near Coulomb impurity, leading to the atomic collapse. Such effects will screen Coulomb impurity much earlier than the actual critical charge of infinite lattice is reached. In the present paper we study how the critical charge of impurity depends on potential localization as well as modified by in-gap edge states.

\begin{figure}
	\includegraphics[scale=0.37]{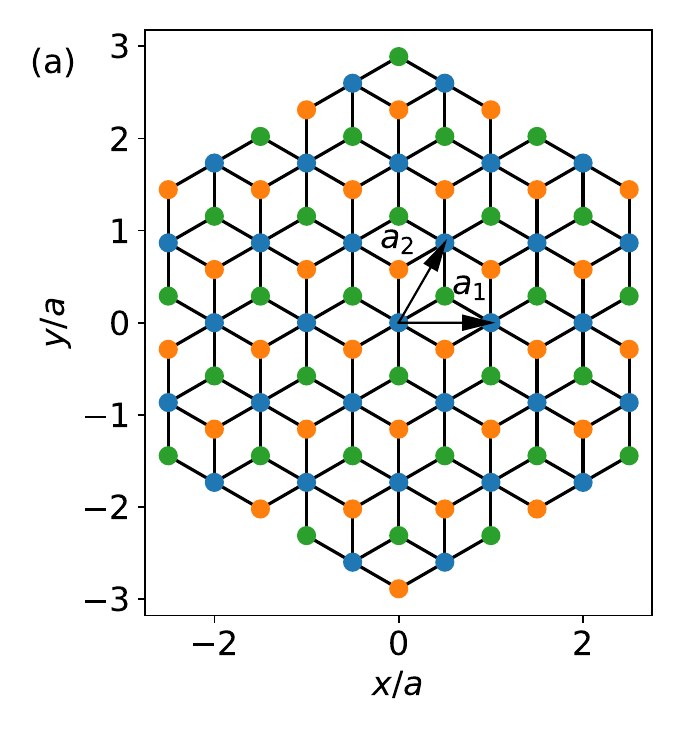}
	\includegraphics[scale=0.37]{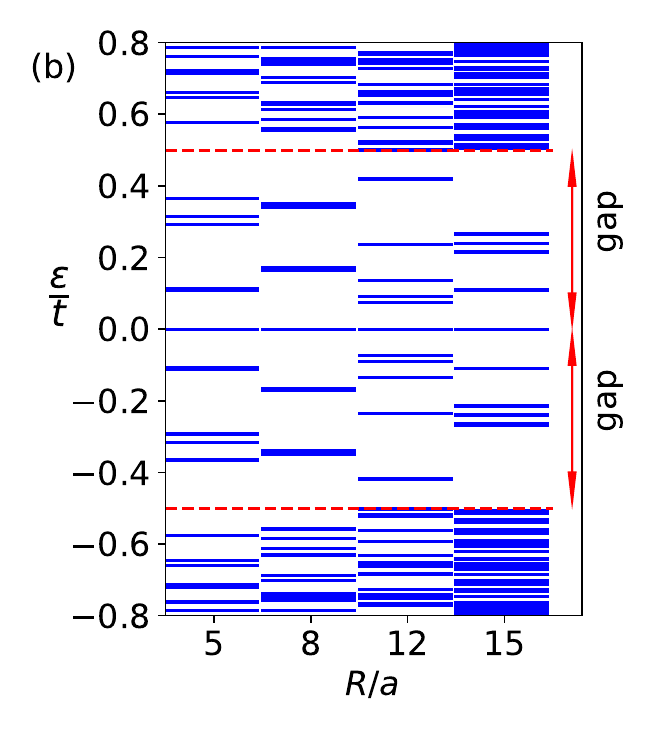}
	\caption{(a) Schematic plot of the dice lattice for quantum dot of radius $R=3a$. The spectrum of quantum dot appearing for several values of radius are shown in (b) for gap parameter $m=0.5 t$. The appearance of in-gap edge states is marked by arrows on the right, that denote gapped regions of the bulk. For the small quantum dot sizes ($R=5a$ and $R=8a$ in the plot) the extension of a spectral gap coming from size quantization is clearly visible.}
	\label{fig:lattice-plot}
\end{figure}
A number of properties of finite size dice lattice quantum dots were studied in recent years. Among them are the description of distributions of edge currents in quantum dots \cite{Soni2020}, prediction appearance of Majorana corner states in the presence of Rashba coupling \cite{Mohanta2022}, size dependence of Landau levels formed in the ring made of $\alpha-T_3$ lattice \cite{Islam2023}, analysis of the role of atomic effects in narrow zigzag ribbons \cite{Hao2023}, valley filtering \cite{Filusch2021,Filusch2022} and dynamical formation of bound states by external driving \cite{Filusch2020} in $\alpha-T_3$  lattice quantum dots.
The study of the effect of edges on atomic collapse in graphene nanoribbons was made in Ref.\cite{Wang2021PRB_graphene}.

The paper is organized as follows: in Sec.\ref{sec:quantum-dots-model} we describe the models of quantum dots used in tight-binding simulations. In Sec.\ref{sec:Coulomb-power-law} we describe the simulation technique and perform the brief analysis of atomic collapse in large quantum dots for the usual and screened Coulomb potential modeled via faster decaying coordinate dependence. In Sec.\ref{sec:Coulomb-edge-states} we analyze the atomic collapse for the same potentials in a small quantum wells. The presence of edge states modifies the results comparing to large systems.
In addition, we study a possibility of creating supercritical impurity by localized potential well in the center of the system. Different positions of such well are analyzed in Sec.\ref{sec:potential-well}. Finally, we compare all results to those obtained for large \cite{Wang2022PRB} and infinite systems \cite{Gorbar2019,Pottelberge2020} in Sec.\ref{sec:conclusions} and make the conclusions about expected observable signatures of atomic collapse in experiments with small artificial lattices.

\section{Quantum dots made of dice lattice}
\label{sec:quantum-dots-model} 
In the present paper we consider a dice lattice as a representative example of flat band model which hosts pseudospin-1 fermions. The previous extensive theoretical studies of atomic collapse in dice lattice allow one to distinguish effects that appear due to finite size from those that was characteristic to infinite model. The geometry is shown in Fig.\ref{fig:lattice-plot}(a). It contains three sublattices, two of which (A,B) are connected only with third one (C) by corresponding hopping parameters $t_{AC}$ and $t_{BC}$. The basis vectors of underlying Bravais lattice are $\vec{a}_1=(1,0)a$ and $\vec{a}_2=(1/2,\sqrt{3}/2)a$.
The tight-binding equations describing this system are \cite{Bercioux2009,Raoux2014,Piechon2015}:
\begin{align}
 \varepsilon \Psi_A(\mathbf{r})=&m_A \Psi_A(\mathbf{r})-t_{AC} \sum_j \Psi_C\left(\mathbf{r}+\boldsymbol{\delta}_j\right),\nonumber\\
 \varepsilon \Psi_B(\mathbf{r})=&m_B \Psi_B(\mathbf{r})-t_{BC} \sum_j \Psi_C\left(\mathbf{r}+\boldsymbol{\delta}_j\right),\\
\varepsilon \Psi_C(\mathbf{r})=&m_C\Psi_C(\mathbf{r})-t_{AC} \sum_j \Psi_A\left(\mathbf{r}+\boldsymbol{\delta}_j\right)\nn
&-t_{BC} \sum_j \Psi_B\left(\mathbf{r}-\boldsymbol{\delta}_j\right)\nonumber.
\end{align}
Here the vectors $\delta_j$ connect neighboring $A$ atoms with $C$ atom and are defined as \cite{Oriekhov2018} $\boldsymbol{\delta}_1=\frac{\mathbf{a}_1+\mathbf{a}_2}{3}$,  $\boldsymbol{\delta}_2=\frac{\mathbf{a}_3+\mathbf{a}_1}{3}$, $ \boldsymbol{\delta}_3=\frac{\mathbf{a}_2+\mathbf{a}_3}{3}$ with $\mathbf{a}_3=\mathbf{a}_2-\mathbf{a}_1$. Below we use the equal hopping parameters $t_AC=t_BC=t$ that correspond to the dice model.
It is possible to open a gap in spectrum and separate the flat band from others by adding onsite potentials of opposite signs to $A$ and $B$ sublattices: $m_A=-m_B=m$ and $m_C=0$. The spectrum of such model consists of three bands separated by gaps, which have the following dispersion in k-space: 
\begin{align}\label{eq:dispersion}
&\epsilon_{\pm}(\vec{k})=\pm\sqrt{m^2+2|f(\vec{k})^2|},\,\,\epsilon_0=0,\\
&f(\vec{k})=-t\left(1+\mathrm{e}^{-i \mathbf{k} \mathbf{a}_2}+\mathrm{e}^{-i \mathbf{k} \mathbf{a}_3}\right)\nonumber.
\end{align}
In addition, one should expect the appearance of in-gap edge states that would be specific for the type of lattice termination chosen. The examples of spectrum for several sizes of circular quantum dot are shown in panel (b) of Fig.\ref{fig:lattice-plot}. The atomic collapse in the infinite lattice with such gap term was studied in Refs.\cite{Gorbar2019,Pottelberge2020}. 
Below we will introduce such gap term to separate the effects produced by in-gap edge states and make them visible in spectral plots.


\subsection{Description of model potentials and tight-binding simulations}
The approach used in the simulations is similar to that described in Refs.\cite{Wang2021PRB_graphene,Wang2022PRB}. We start with the lattice model of quantum dot implement with the help of Kwant code \cite{Groth2014kwant}. The spectrum of finite size system contains all levels coming from bulk and edge states. They evolve differently with growing charge of impurity, in particular the far-away placed edge states stay nearly at the same energy. In order to avoid complicated procedures of distinguishing atomic collapse states from such edge states, we calculate a energy-resolved local density of states (LDOS) in the central unit cell of the quantum dot. The calculation is performed by using kernel polynomial method with a high number of moments (typically of the order of $10^4$) that ensures precise energy resolution of the plots \cite{Weisse2006RevModPhys}. 
The calculation of LDOS also guarantees that the shown states are only those that contribute into physics in the vicinity of impurity and are possible to measure via STM-type techniques.

The model potentials used to describe a normal and screened Coulomb impurity made as adatom on top of the lattice are:
\begin{align}\label{eq:potential_models}
	V(\vec{r})=-V\frac{a^n}{(|\vec{r}|^2+r_0^2)^{n/2}}.
\end{align}
\begin{figure*}
	\centering
	$\text{(a)}$\includegraphics[scale=0.7]{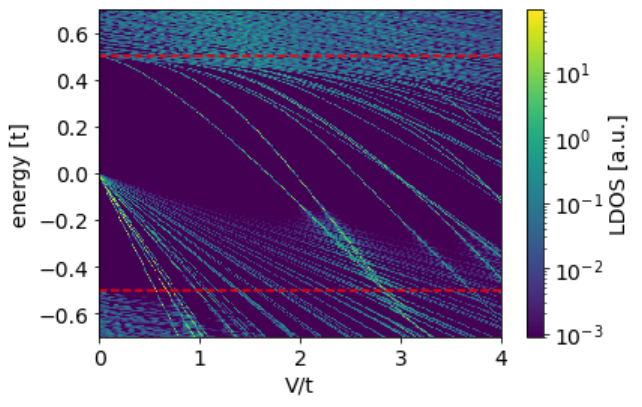}
	$\text{(b)}$\includegraphics[scale=0.7]{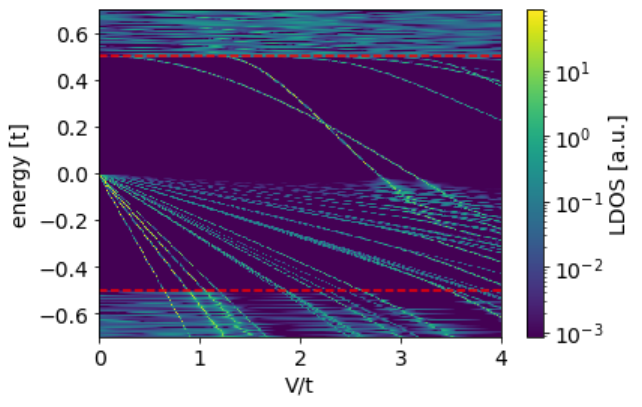}
	$\text{(c)}$\includegraphics[scale=0.7]{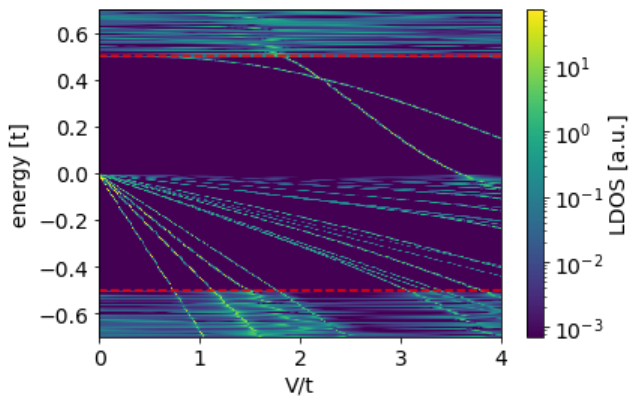}
	$\text{(d)}$\includegraphics[scale=0.7]{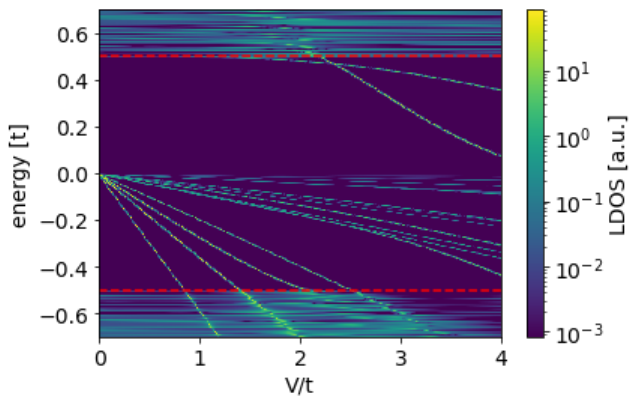}
	\caption{LDOS in central unit cell as a function of potential strength for power $n=1, 2$ upper row and $n=3,4$ lower row. Quantum dot size is $R=50 a$. Regularization radius was taken $r_0= a$. The gap size it $m=0.5 t$. The (a) $n=1$ power corresponds to bare Coulomb potential and demonstrates the largest density of bound states decoupled from flat band.}
	\label{fig:potentials-r-50}
\end{figure*}
Here $r_0$ is the regularization radius of impurity and $n=1$ corresponds to the pure Coulomb potential. Below in the numerical calculations we fix the depth of potentials with different $n$ to be the same at $\vec{r}=0$. This is done by setting $r_0=a$ for all potentials. Another choice would be to set different $r_0$ depending on powers $n$ and keep the constant $(a/r_0)^n$ the same. That would result in effective normalization of $V$ parameter and produce qualitatively the same results. As a motivation for such model of long- and short-range potentials we note that the $n=3$ power case describes the Coulomb impurity with dynamical one-loop polarization screening in monolayer graphene \cite{Katsnelson2006a}. The comparison of bound state picture that decouple from flat band in such potentials in a lattice model would highlight the effect of localized structure of a flat band states. Indeed, to prove that flat band decomposes into continuum-type spectrum in Coulomb potential, a special convergence-checking technique was used in effective low-energy model \cite{Pottelberge2020}. Below we show that lattice calculations support that conclusion with correction to finite state number in the system, while the short-range potentials demonstrate different behavior.

\section{Atomic collapse in Coulomb and power-law potentials}
\label{sec:Coulomb-power-law}
In this section we analyze how the bound states evolve with potential strength $V$ of impurity for a large system size. Specifically, we take the quantum dot of radius $R=50 a$ centered at $C$ atom and place impurity at the center of a disk. The results of simulation are shown in Fig.\ref{fig:potentials-r-50} for powers $n=1$ to $n=4$ of model potential \eqref{eq:potential_models}.
The results presented in this section for a Coulomb potential are in agreement with results of Ref.\cite{Wang2022PRB}. In addition we tested the parameters of Ref.\cite{Wang2022PRB} to check the convergence of the simulations. In the next section we analyze the smaller system sizes to uncover effects of edge states on atomic collapse in quantum dots. 

From the spectral plots shown in Fig.\ref{fig:potentials-r-50} the following conclusions can be made: the critical charge for the first bound state decoupled from the upper band is larger than for the second only for Coulomb potential with $n=1$. For the infinite system such structure in spectrum appeared for states with angular momentum $j=0$ and $j=1$ with $j=1$ firstly decoupling from continuum \cite{Gorbar2019}. For the large system size the in-gap states do not influence atomic collapse happening in the bulk. 

For the flat band the density of bound states that decouple at small charges becomes lower with increasing power of potential $n$. The energy separation of such states also becomes larger with $n$. This agrees with the prediction of Refs.\cite{Gorbar2019,Pottelberge2020} that flat band decomposes into continuum of states in the long-range potential, but only a discrete set of bound states decouples in short-range potential. The localized structure of flat band states leads to approximately linear behavior with growing charge parameter of impurity $V$. The slope is defined by the distance from the localization point to the potential center. 

\section{Atomic collapse in small quantum dots: the role of edge states}
\label{sec:Coulomb-edge-states}
 \begin{figure*}
	\centering
	$\text{(a)}$\includegraphics[scale=0.7]{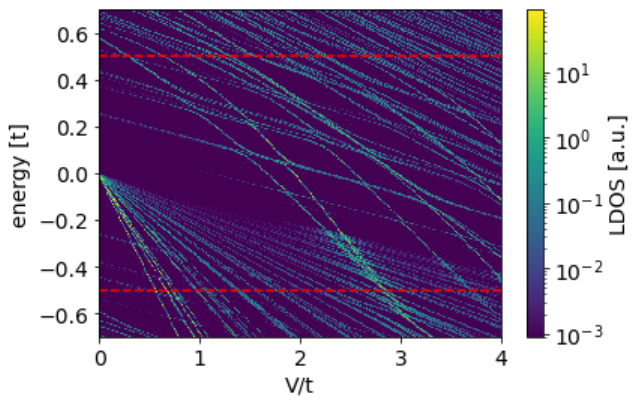}\quad
	$\text{(b)}$\includegraphics[scale=0.7]{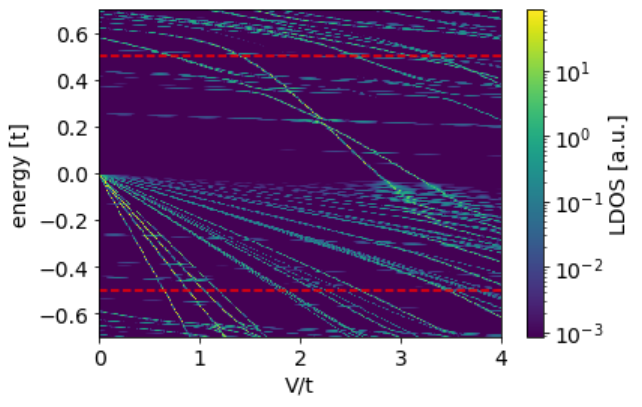}
	$\text{(c)}$\includegraphics[scale=0.7]{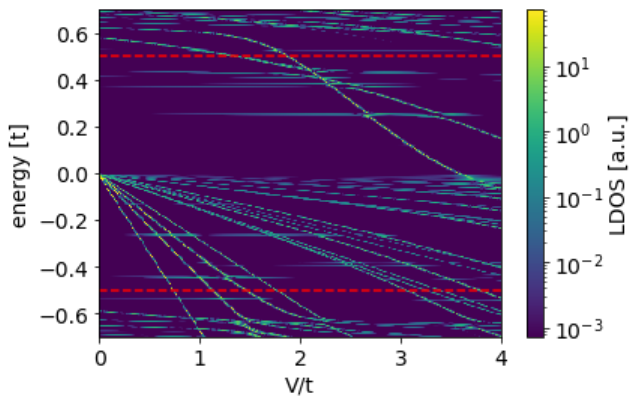}\quad
	$\text{(d)}$\includegraphics[scale=0.7]{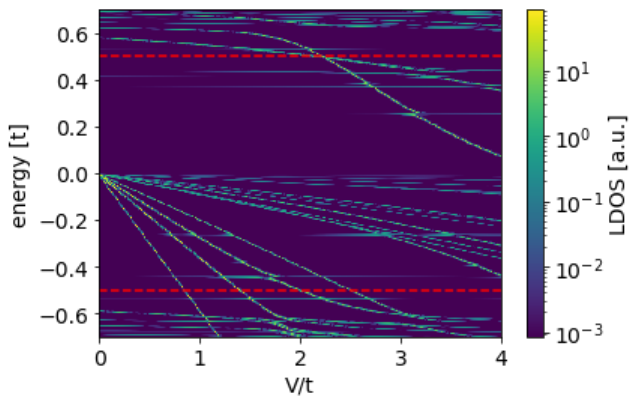}
	\caption{LDOS in central unit cell as a function of potential strength for $n=1, 2$ upper row and $n=3,4$ lower row. Disk size is $R=10a$. Note that these results are for intermediate disk size, which would describe the artificial lattices made of a set of individual sites (atoms or clusters). Regularization radius was taken $r_0=1 a$. The gap size it $m=0.5 t$. A number of level repulsions with in-gap edge states are clearly visible in all plots. Several edge states also decrease in energy in a bare Coulomb potential in panel (a).}
	\label{fig:potentials-r-10}
\end{figure*}
In this section we perform the simulations for much smaller system sizes where the edge states contribute to the LDOS at the central unit cell. As was shown in Fig.\ref{fig:lattice-plot}, the possible energies of in-gap edge states strongly depend on the size. We analyze a set of different radius values to describe the qualitative role of edge states on critical charge value. One should note that to obtain precise numerical values a simulation should be made for a sample geometry available in experiment. However, a number of conclusions can be made from analysis of a set of different samples.

Firstly we present the results of calculations for the system with radius $R=10a$, that are shown in Fig.\ref{fig:potentials-r-10} for the powers $n=1$ to $n=4$ of model potential \eqref{eq:potential_models}. Due to the small size of the system all levels in spectrum decrease in energy with increasing charge of impurity. The in-gap edge states demonstrate nearly linear dependence of energy on potential strength $V$, which follows from their localization and nearly zero kinetic energy. But, the resonant anticrossing of bound states decoupled from the bands with such in-gap states leand to the appearance of nontrivial dependence on $V$ for both. In addition, such anticrossings modify the value of critical charge for the dive-into-continuum type problem. We define such charge as the position of the first intersection between levels originating from different bands in infinite model (e.g., upper band levels to flat band, and flat-to-lower band).
In the case when in-gap states are filled, one may expect much lower values of critical charge required to screen the impurity by electron tunneled to bound state.

\subsection{Size effect on a critical charge value}
\begin{figure}
	\centering
	\includegraphics[scale=0.5]{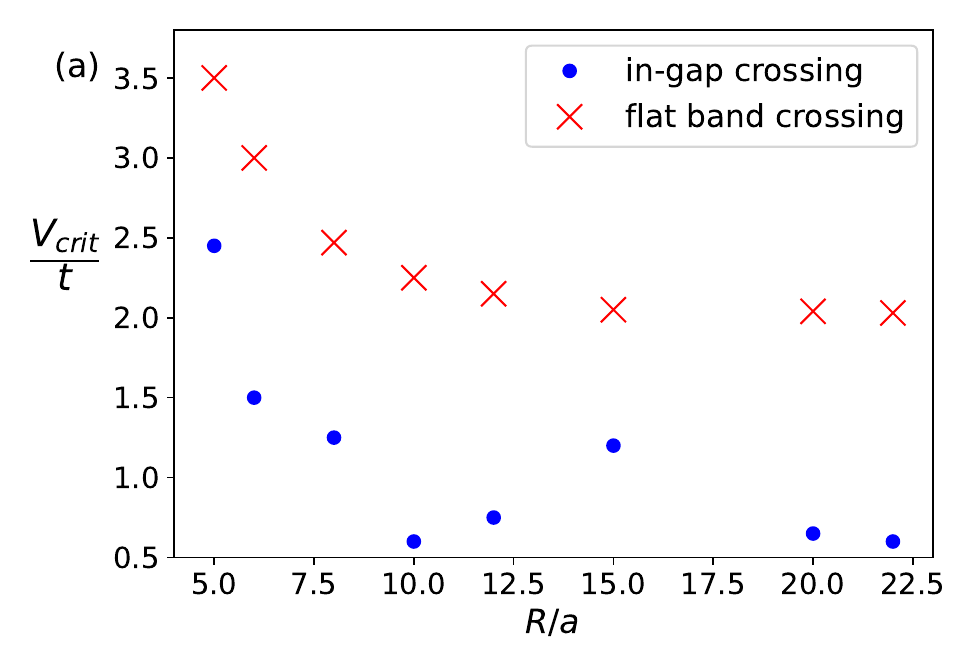}
	\includegraphics[scale=0.5]{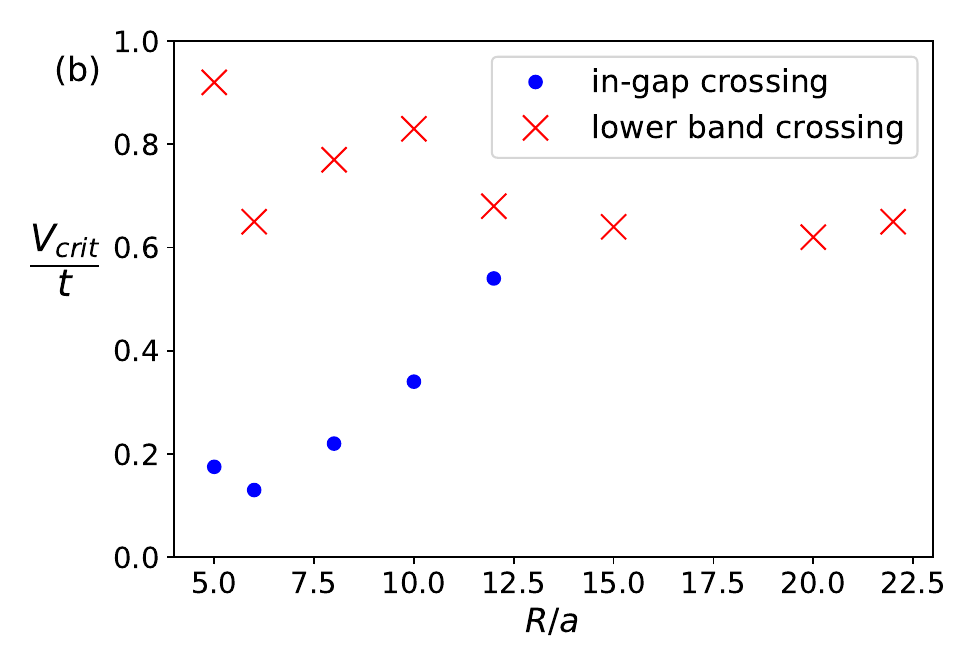}
	\caption{Size dependence of potential values at which the first repulsion with bound in-gap state happens and when the dive-into-continuum phenomenon between states originating from different bands is happening. Panel (a) shows the results for states originating from upper band, and panel (b) for the states from flat band. }
	\label{fig:crit-charges-size-dep}
\end{figure}

Next we analyze the role of radius of quantum dot on the critical charge values as well as first level repulsion with in-gap state in Coulomb potential.
The Fig.\ref{fig:crit-charges-size-dep} shows results for the levels decoupled from upper band in panel (a) and from the flat band in panel (b). Both plots show that the critical charge for dive-into-continuum charge grows with descreasing size. However, in addition to the effect of trivial gap size growing due to  the discretization of momentum
\begin{align}
	\delta_m \sim \frac{t^2}{m} \left(\frac{\pi a}{R}\right)^2,
\end{align}
that follows from linearized version of dispersion \eqref{eq:dispersion}, two other effects play a role. As was discussed above, in small-scale finite system all levels decrease in energy in external long-range potential.
In addition, the nontrivial role is played by the presence of one or several resonant anti-crossings with in-gap bound states, that charge the critical charge value. Thus, for the flat band we observe non-monotonous behavior of dive-into-continuum $V_{crit}$ values. 

\section{Supercritical localized potential well}
\label{sec:potential-well}
In this section we analyze the possibility of forming the dive-into-continuum bound states by a localized potential well. As an example of experimental implementation one can consider setup from Ref.\cite{Jiang2017NNano} where the atomic collapse in graphene was observed in potential well of circular p–n junction. We  investigate the potential wells that are centered at single sublattice. Other potential wells that cover several sites are very close to model potentials \eqref{eq:potential_models} with large $n$ (already $n=4$ is close to localized potential well) and would not lead to qualitatively new results. 

\begin{figure*}
	\centering
	\includegraphics[scale=0.55]{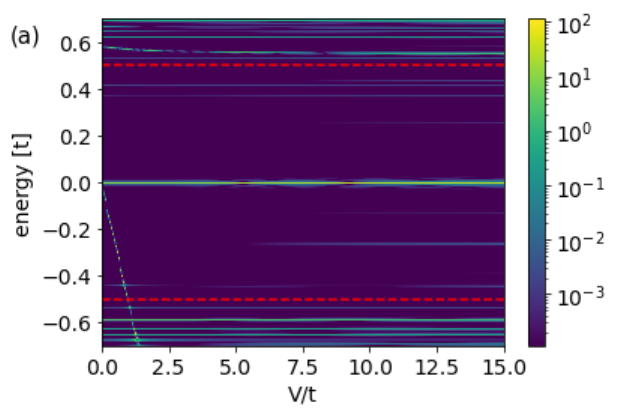}
	\includegraphics[scale=0.55]{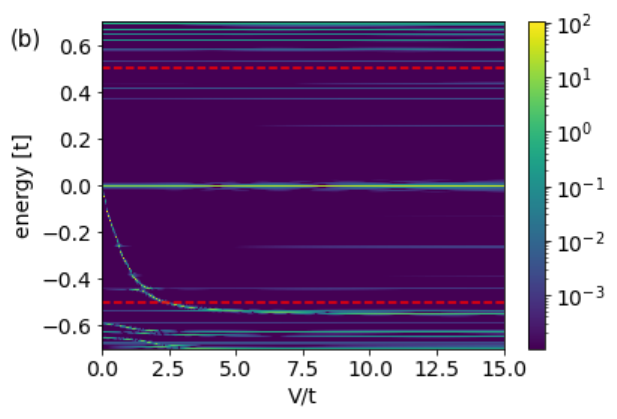}
	\includegraphics[scale=0.55]{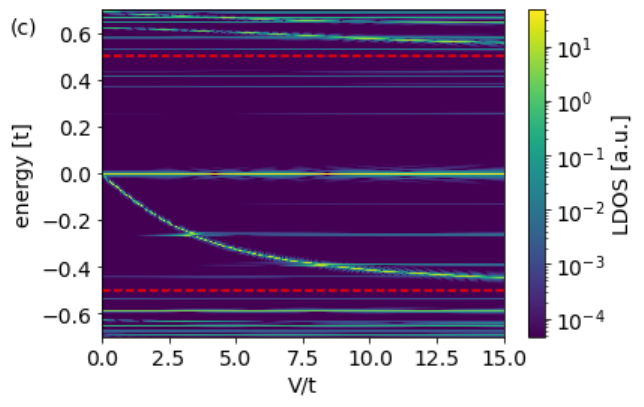}
	\includegraphics[scale=0.6]{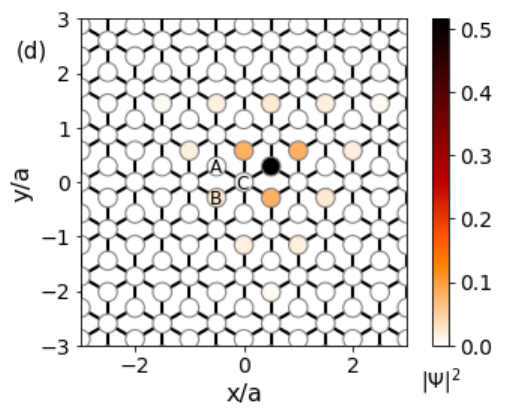}\quad\quad
	\includegraphics[scale=0.6]{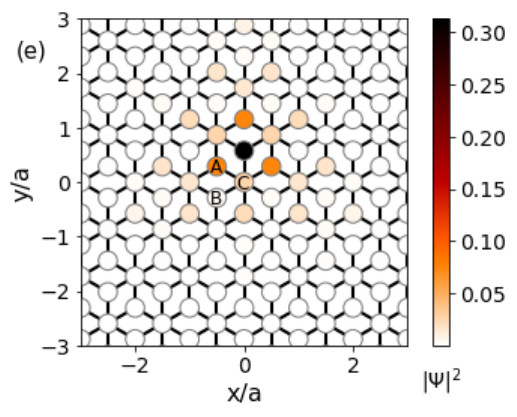}\quad\quad
	\includegraphics[scale=0.6]{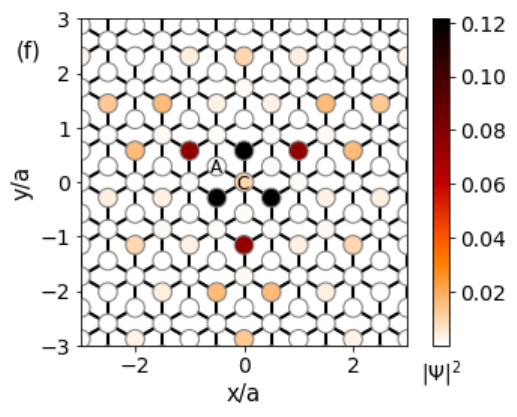}
	\caption{Upper row: LDOS in central unit cell as a function of potential strength for potential well localized  at single site (a) $A$, (b) $B$ and (c) $C$ respectively. Quantum dot size is $R=10a$ and the gap size is $m=0.5 t$. The bound state for $C$-localized well asymptotically reaches the lower continuum. Lower row: wave function distribution for a bound state at potential values (d), (e) $V=0.6t$, and (f) $V=10 t$ with localizations at $A$, $B$ and $C$ corresponding to upper row.}
	\label{fig:potential-well-single-r-10}
\end{figure*}

The results of calculation of LDOS are presented in upper row of Fig.\ref{fig:potential-well-single-r-10}. The localized potential well decouples bound states from the flat band much earlier than from upper band. In the plots shown in Fig.\ref{fig:potential-well-single-r-10} we focus on the evolution of energy for such bound states with increasing charge. Only single state is decoupled as there is only single state in flat band localized exactly at the position of potential well. Notably, for the $C$-localized potential well the bound state approaches the lower band asymptotically with growing $V$. Also this bound state has the largest value of occurred level repulsions inside the gap, which leads to the largest delocalization of its wave function. 

\begin{figure}
	\centering
	\includegraphics[scale=0.5]{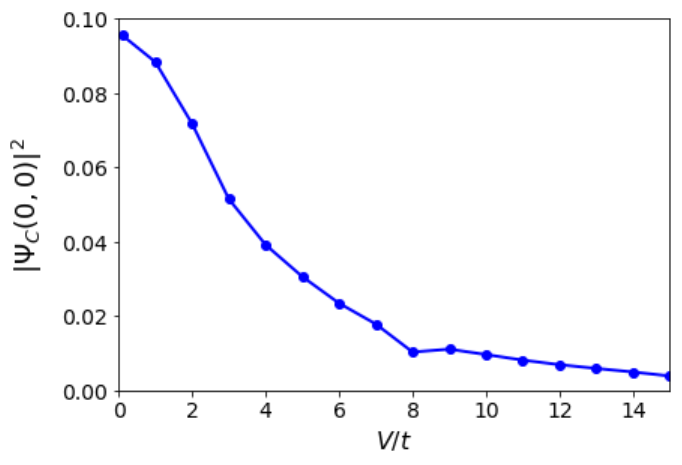}
	\caption{Dependence of C-component of wave function density $|\Psi_C(\vec{r}=0)|^2$ on the central atom for the potential well localized at $C$-atom. The decreasing value of wave function compensates the growing potential value, leading to the asymptotic approaching of the lower continuum by a bound state decoupled from a flat band.}
	\label{fig:C-potential-C-dep}
\end{figure}

The lower row in Fig.\ref{fig:potential-well-single-r-10} shows the distribution of wave function for a bound state at such values of potential that the energy level is placed inside the gap. For all cases the distribution shows $C_3$-symmetric pattern keeping the discrete symmetry of the lattice. The notable difference between $A$-, $B$-localized and $C$-localized potential wells is that in the first two cases the highest density is placed exactly at potential well site. Instead, for the $C$-localized well the maximum density values are distributed symmetrically around $C$ site. The state itself is not as localized as two other examples, and has a small density at the site with potential. Indeed, by calculating density $|\Psi_C(\vec{r}=0)|^2$ at central $C$ atom with potential well for different values of $V$, we find that it decreases with growing $V$ (see Fig.\ref{fig:C-potential-C-dep}). As was shown in Ref.\cite{Gorbar2019}, the $C$-component of a wave function in a free continuum model behaves as $\Psi_C\sim (\epsilon^2-m^2)$, tending to zero when $\epsilon\to \pm m$. Such dependence on energy compensates the potential strength and leads to asymptotic approaching of lower band by a bound state (see Fig.\ref{fig:potential-well-single-r-10}(c)). The small dip at $V=8 t$ in Fig.\ref{fig:C-potential-C-dep} is related to the repulsion with in-gap energy level and does not affect the general qualitative picture.

In addition, we note that the bound states decoupled from the flat band for the $A$- and $B$-localized potential demonstrate near linear behavior of energy dependence with growing potential. The level repulsion with in-gap state for $B$-localized quantum well leads to the larger critical charge. The asymmetry between $A$- and $B$-localized well cases appears due to the gap that takes $+m$ value at $A$ site and $-m$ at $B$ site, while potential has the same sign $-V$. 
 
\section{Conclusions}
\label{sec:conclusions}
In the present paper we analyzed the role of size effect on the possibility of observing atomic collapse phenomenon in the quantum dots made of dice lattice. Depending on whether in-gap edge states are filled, the critical charge takes different values for the same system size. The size dependence itself of critical charge for dive-into-continuum problem demonstrates non-monotonous behavior due to a number of level-repulsion events happening with the bound state inside the gap. These level repulsions in spectrum show the appearance of superposition-type states, localized partially on the interior of the system, where the potential is weak. That makes the effect of potential weaker and requires larger charge to put bound state energy below the gap value.   

In addition to the size effect happening due to finite radius of quantum dot, we analyzed the role of the effective range of potential in decoupling bound states from flat band. It is shown that while the long-range potential decomposes flat band into a dense set of bound states, more short-range potentials result in large gap separations between lowest bound states. Such effects might occur when the Coulomb impurity is dynamically screened. In addition, this result supports the conclusion about full decomposition of a flat band by long-range potential, found earlier in a continuum model by solving problem exactly at zero energy \cite{Gorbar2019} and showing the convergence to continuum spectrum of bound states numerically \cite{Pottelberge2020}.

Next we analyzed the possibility of creating supercritical localized potential well on a single site. The structure of wave functions in the dice model allows the formation of bound state decoupled from the flat band, but makes the threshold potential very large for the decoupling of bound state from the upper band. The dive-into-continuum situation occurs only to rim-placed potential wells, but not for hub-centered. This can be explained by the fact that wave functions has zero hub component exactly at the gap edge, thus making it impossible to cross by a bound state localized only on a hub site. In conclusion we note that these results describe the potential possibility of finding atomic collapse in artificial electronic flat band lattices, where the single-particle physics might be easier to study due to the larger lattice constant \cite{Slot2017} and weaker electron-electron interactions.

\begin{acknowledgements}
	We are grateful to E. V. Gorbar and V. P. Gusynin for fruitful discussions.  D. O. O. acknowledges the support from the Netherlands Organization for Scientific Research (NWO/OCW) and from the European Research Council (ERC) under the European Union's Horizon 2020 research and innovation program.
\end{acknowledgements}
\bibliography{coulomb_finite_lattice_bib}
\end{document}